\documentclass[prl,nofootinbib,twocolumn,preprintnumbers,superscriptaddress]{revtex4}
\usepackage[dvips]{graphicx}
\usepackage{amsmath,amssymb,dsfont}
\newcommand{\beq}{\begin{eqnarray}}
\newcommand{\eeq}{\end{eqnarray}}

\def\ltap{\ \raise.3ex\hbox{$<$\kern-.75em\lower1ex\hbox{$\sim$}}\ }
\def\gtap{\ \raise.3ex\hbox{$>$\kern-.75em\lower1ex\hbox{$\sim$}}\ }

\def\eg{{\it e.g.}}

\def\be{\begin{equation}}
\def\ee{\end{equation}}
\def\bea{\begin{eqnarray}}
\def\eea{\end{eqnarray}}

\def\mysection#1{{{\bf #1}.~}}
\newcommand{\vckm}{V^{\rm CKM}}
\newcommand{\mkk}{m_{\rm KK}}
\newcommand{\ttb}{t \bar t}
\newcommand{\ecm}{e\, \mathrm{cm}}
\begin{document}

\title{Extraordinary Phenomenology from Warped Flavor Triviality}
\author{C\'edric Delaunay}
\affiliation{Department of Particle Physics, Weizmann Institute
of Science, Rehovot 76100, Israel}
\author{Oram Gedalia}
\affiliation{Department of Particle Physics, Weizmann Institute
of Science, Rehovot 76100, Israel}
\author{Seung J. Lee}
\affiliation{Department of Particle Physics, Weizmann Institute
of Science, Rehovot 76100, Israel}
\author{Gilad Perez}
\affiliation{Department of Particle Physics, Weizmann Institute
of Science, Rehovot 76100, Israel}
\author{Eduardo Pont\'on }
\affiliation{Department of Physics, Columbia University, 538 W.
120th St., New York, NY 10027, USA}

\begin{abstract}
Anarchic warped extra dimensional models provide a solution to
the hierarchy problem. They can also account for the observed
flavor hierarchies, but only at the expense of little hierarchy
and CP problems, which naturally require a Kaluza-Klein (KK)
scale beyond the LHC reach. We have recently shown that when
flavor issues are decoupled, and assumed to be solved by UV
physics, the framework's parameter space greatly opens. Given
the possibility of a lower KK scale and composite light quarks,
this class of flavor triviality models enjoys a rather
exceptional phenomenology, which is the focus of this letter.
We also revisit the anarchic RS EDM problem, which requires
$m_{\rm KK}\gtrsim 12\,$TeV, and show that it is solved within
flavor triviality models. Interestingly, our framework can
induce a sizable differential $t\bar t$ forward-backward
asymmetry, and leads to an excess of massive boosted di-jet
events, which may be linked to the recent findings of the CDF
Collaboration. This feature may be observed by looking at the
corresponding planar flow distribution, which is presented
here. Finally we point out that the celebrated standard model
preference towards a light Higgs is significantly reduced
within our framework.
\end{abstract}

\maketitle

\mysection{Introduction} The Randall-Sundrum (RS) warped extra
dimensional framework provides a solution to the hierarchy
problem~\cite{Randall:1999ee}. The most studied version of this
class of models is the ``anarchic bulk RS'' scenario, where the
standard model (SM) fields propagate in the 5D bulk, and the
microscopic flavor parameters are generic. The SM gauge group
is enlarged to contain a product of $SU(2)$ and discrete
custodial symmetries~\cite{Agashe:2003zs, Agashe:2006at}, thus
greatly suppressing RS corrections to the electroweak (EW)
observables. However, a closer look at this scenario shows
that, despite providing a solution to the SM flavor
puzzle~\cite{Huber:2003tu}, little hierarchies and CP problems
remain, pushing the Kaluza-Klein (KK) scale of these models up
to unnatural values. For instance, the contribution to electric
dipole moments (EDMs) was found to be very roughly $20$ times
above the present bounds for an ${\cal O}(3\,\rm TeV)$ KK
scale~\cite{aps}, the combined contribution to
$\epsilon_K$~\cite{Davidson:2007si,Csaki:2008zd,Bona:2007vi}
and $\epsilon'/\epsilon_K$~\cite{epsp} requires a KK scale of
${\cal O}(6\,\rm TeV)$, and a careful EW fit to a subclass of
these models (at the one loop level)~\cite{Delaunay:2010dw}
recently showed that a KK scale above ${\cal O}(4\,\rm TeV)$ is
required. All this motivates trying to naturally decouple the
flavor and CP issues, which have nothing to do with the EW
fine-tuning problem, and analyzing the status of this framework
when the EW scale naturalness is the main concern. This was the
subject of~\cite{Delaunay:2010dw}, which studied the ``flavor
trivial'' case where the flavor hierarchy is set by UV physics
on the Planck brane~\cite{Rattazzi:2000hs}. The resulting bulk
RS phenomenology turns out to be significantly modified, with a
greatly improved EW fit allowing for a KK mass below
2\,TeV~\cite{Delaunay:2010dw}, while the RS $\epsilon_K$
problem is absent.

The purpose of the present letter is to demonstrate that this
class of flavor triviality models leads to exceptional collider
phenomenology. We show that the same features that lead to a
successful fit to the EW and flavor-violating observables, also
yield unconventional signals. In fact, some of the observables
discussed below have been already measured at the Tevatron in
the context of the forward-backward asymmetry (FBA) in top pair
production, a search for highly boosted tops and a study of the
planar flow distribution of massive
jets~\cite{cdfttbarfb,D0ttbarfb,cdfttbarfbnew,boostedtops,
substructure}. It is rather intriguing that some
inconsistencies with the SM predictions have been observed. If
confirmed at the LHC, they may support the presence of some
sort of flavor triviality (for a complimentary study of
multi-tops at the LHC see~\cite{Jung:2010ms}). We also discuss
how the bound on the Higgs mass is typically softened in our
framework, and provide a quantitative analysis of the RS EDM
problem, showing that it is naturally solved with flavor
triviality.

\mysection{The model} We work in a slice of AdS$_5$ space-time,
whose fifth (conformal) coordinate $z$ is bounded by two
branes, at $R=M_{\rm \overline{Pl}}^{-1}\sim
(10^{19}\,$GeV$)^{-1}$ in the UV and $R\,'\sim\,$TeV$^{-1}$ in
the IR, where $M_{\rm \overline{Pl}}$ is the reduced Planck
mass. We use the notation $\epsilon\equiv e^{-\xi}$, where
$\xi\equiv\log(R'/R)$. We impose a $SU(2)_L \times SU(2)_R
\times U(1)_X$ gauge symmetry in the bulk, and assume that the
Higgs field, H, is a bulk field with vacuum expectation value
(VEV) $\langle H\rangle= vR'/R^{3/2}
\sqrt{1+\beta}(z/R')^{2+\beta}$ with $v\simeq 246\,$GeV. The
VEV localization in the bulk is set by $\beta$, and $\beta=0$
corresponds to gauge-Higgs unified models. The SM fermions are
embedded as $Q\sim(2,2)_{2/3}$, $U\sim(1,1)_{2/3}$,
$D\sim(1,3)_{2/3}\oplus (3,1)_{2/3}$ and $L\sim(2,2)_0$,
$E\sim(1,3)_{0}\oplus(3,1)_{0}\,$.

We also gauge in the bulk the non-abelian part of the SM flavor
symmetry $SU(3)_Q \times SU(3)_U \times SU(3)_D \times SU(3)_L
\times SU(3)_E$. The breaking of the flavor group occurs on the
UV brane, and is shined towards the IR by some flavon scalar
fields, $\Phi$, whose VEVs are proportional to the 5D Yukawa
matrices $Y_{U,D,E}$. Thus, in contrast to previous scenarios,
we take the 5D Yukawas to be hierarchical (similar to the 4D
picture), and set by unspecified UV physics. All flavor
changing effects are then controlled by the SM Yukawa
couplings, thus realizing the minimal flavor violation (MFV)
ansatz.

Within this model, it is possible to find sweet spots in the
parameter space of the fermion bulk masses described by
$c_{Q^3}$, $c_t$, $c_b$, $c_{Q^i}$, $c_{U^i}$ and $c_{D^i}$
($i=1,2$, with universal first two generation masses) for the
quark sector and $c_L$ and $c_E$ for the leptons (taken to be
fully universal).

\mysection{Sweet Spot} In order to quantify the
phenomenological aspects analyzed below, we consider a specific
set of parameters close to one of the sweet spots presented
in~\cite{Delaunay:2010dw}:
\beq \label{sweetspot}
C_Q &\simeq& (0.50, 0.50, 0.02)\,, \ \ C_D \simeq (0.63, 0.63,
0.57)\,, \nonumber \\ C_U &\simeq& (0.15, 0.15, 0.48)\,.
\eeq
The related effective 5D Yukawa eigenvalues (which are
accompanied by $\alpha_{U,D}$ when coupled to the Higgs, \eg\
in the fermion masses) are:
\be \label{sweetspotyuk}
\alpha_U Y_U \!\simeq\! (4.3 \times 10^{-5}, 0.021, 4.2)\,, \,
\alpha_D Y_D \!\simeq\! (0.01, 0.19, 0.45).
\ee
The corresponding $2\sigma$ EW bound on the KK scale is
$\mkk\gtrsim 1.7\,(2.1)$~TeV for a six (one) parameter fit. In
order to make this sweet spot consistent with flavor bounds, we
can choose $\alpha_{U,D}$ such that the 5D bottom Yukawa
becomes much bigger than the top one, \eg\ $\alpha_{U,D}\simeq
4, 0.12$. This leads to down alignment: $[m_D,Y_D] \simeq 0$,
which effectively removes all constraints coming from the down
sector. Consequently, the above set of parameters complies with
flavor, at the cost of a (natural) hierarchy of
$\mathcal{O}(30)$ between the $\alpha$'s. Note that since the
one-loop contribution to $\delta g_{Z\bar{b}b}$ at large
$\alpha_D Y_D$ is not known yet~\cite{ZbbYb}, we conservatively
consider only cases with $(\alpha_D Y_D)^2\ll (\alpha_U
Y_U)^2$~\cite{Delaunay:2010dw}. Relaxing this constraint would
probably lead to a larger set of viable models where no
significant hierarchy between the $\alpha$'s is required.

\mysection{Electric Dipole Moments} We now analyze the
constraints from null EDM searches. We consider the flavor
triviality case and also derive a robust bound for the anarchic
class of models. To date, the strongest bounds come from
neutron, Mercury and Thallium EDM searches. We find that the
neutron and Mercury EDMs~\cite{Baker:2006ts,Griffith:2009zz},
\beq
\begin{split}
|d_n^{\,\rm exp}|&\lesssim 2.9 \times 10^{-26}\,e\,{\rm cm}
\quad (95\%~\textrm{CL}), \\ \left|d_{Hg}^{\,\rm exp}\right|
&\lesssim 3.1 \times 10^{-29} \,e\,{\rm cm} \quad
(95\%~\textrm{CL}),
\end{split}
\eeq
are of most relevance in our case. These observables are
measured far below the QCD scale.

The neutron EDM is sensitive to the CP-odd dipole effective
operator at the TeV scale $ 2i\times\mathcal{L}_{\rm
eff}\supset \sum_f d_f^E \bar{f} \left(\sigma\cdot F\right)
\gamma_5 f \,,$ where $f=d,s,e$ etc.\ stands for a fermion
flavor and $F$ is the photon field strength. We use the parton
quark model (PQM) to relate the neutron EDM to this set of
operators, since it is the only nuclear model including the
strange quark contribution, which turns out to dominate in RS.
The neutron EDM is then given by (see \eg~\cite{reloaded} and
refs.~therein)
\be
d_n=\eta^E \left( \Delta^{\rm PQM}_d d_d^E +\Delta^{\rm PQM}_u
d_u^E+\Delta^{\rm PQM}_s d_s^E \right)\,,
\ee
where
\beq
\Delta^{\rm PQM}_{d,u,s}\, \simeq \, 0.75,\,-0.51, \,-0.23\,;\
\ \, \eta^E\simeq 1.5 \,,
\eeq
and $d_f^E$ are evaluated at the EW scale.

The mercury EDM is sensitive to several types of operators, but
in the current work the leading contribution is from the
chromo-electric dipole $ 2i\times\mathcal{L}_{\rm eff}\supset
\sum_f d_f^C \bar{f} \left(\sigma\cdot G\right) \gamma_5 f \,$,
where $G$ is the gluon field strength. The relation
is~\cite{reloaded}
\beq \label{hgedm}
d_{Hg}=7 \times 10^{-3} e(d_u^C-d_d^C)/g_s \,,
\eeq
where $g_s$ is the QCD coupling and $d^C_f$ are evaluated at
1~GeV.

Within RS, the dipole operator $d_f^E$ is induced by a one-loop
process with KK-quarks and a Higgs or a KK-gluon~\cite{aps}.
Since to leading order, the KK-gluon exchange diagram is
proportional to $m_D$, hence real, the Higgs contributions are
expected to dominate. The relevant RS amplitude has been
calculated \eg\ in~\cite{epsp}, and the result is
\be \label{fedm}
d_f^E \simeq \frac{ev}{16 \pi^2 \mkk^2} \times ({\rm
spurion})_f \,.
\ee
The chromo-electric dipole is given by Eq.~\eqref{fedm}, with a
proper replacement of the electromagnetic coupling of the quark
with its QCD coupling. Generically, the leading contributions
have the following spurion dependence~\cite{aps}:
\bea \label{dipole_spurions}
\hspace*{-.10cm}({\rm spurion})_{d,s} \hspace*{-.10cm}&=&
\hspace*{-.10cm}\left[ F_Q^\dagger \left(a_N Y_D
Y_D^\dagger+a_CY_U Y_U^\dagger\right) Y_D F_D\right]_{11,22} ,
\nonumber \\ ({\rm spurion})_u \hspace*{-.10cm}&=&
\hspace*{-.10cm} \left[ F_Q^\dagger \left(a_N Y_U
Y_U^\dagger+a_CY_D Y_D^\dagger\right) Y_U F_U \right]_{11} ,
\eea
where $a_{N(C)} $ corresponds to the neutral (charged) Higgs
exchanges, and $F_X$ are spurion matrices whose eigenvalues
$f_{x^i}$ represent the IR projection of the quark zero-mode
profiles: $f_{x^i}^2=(1-2c_{x^i})/( 1-\epsilon^{
1-2c_{x^i}})\,$. Note that for models with a bulk Higgs,
corrections for its overlap with the zero-mode fermions should
be taken into account in Eq.~\eqref{dipole_spurions}, as we do
implicitly throughout the paper.

Below we provide the first robust quantitative bound on the
anarchic case, for which we follow the approach
of~\cite{epsp,Agashe:2008uz}. We look for the weakest bound
which simultaneously minimizes the contributions from
$\epsilon_K\propto \left(Y^*_D\right)^{-2}$, where $Y^*_D$ is
the average value characterizing the anarchic 5D down Yukawa
matrix, and $d_f\propto\left(Y^*_D\right)^2$. We find that the
strongest EDM bound comes from Mercury via $d_d^C\,$.
Conservatively, we focus only on the neutral Higgs exchange,
which amounts to setting $a_C\to 0$ in
Eq.~\eqref{dipole_spurions} (since the charged Higgs
contribution is proportional to $Y_U Y_U^\dagger$, it cannot be
naively added to the neutral one or combined with
$\epsilon_K$). The corresponding one-loop contribution was
calculated in~\cite{epsp,Agashe:2008uz}
\be \label{cedm_anar}
d^C_d \sim \frac{3g_s m_d}{16 \pi^2 \mkk^2} \,
\left(Y^*_D\right)^2 \,,
\ee
yielding\footnote{The overlap correction, which is implicitly
included, is $\sim$0.1~\cite{epsp}.}
\be \label{anar_hg}
d_{Hg} \sim 1.2 \times 10^{-27} \left(Y^*_D\right)^2 \,
\left({\rm TeV}\over \mkk\right)^2 \ecm \,.
\ee
Consequently, the resulting bound on the KK scale is
\be \label{anaredm}
\mkk\gtrsim6.2 \, Y^*_D \textrm{ TeV}\,.
\ee
Optimizing the bound in Eq.~\eqref{anaredm}, w.r.t.~$Y^*_D$,
together with the $\epsilon_K$ bound: $\mkk\gtrsim8.5\,
g_{s*}/Y^*_D$~TeV for a bulk Higgs with $\beta=0$~\cite{epsp}
(where $g_{s*} \approx 3$ is the KK-gluon coupling, including
one loop matching~\cite{Agashe:2008uz}), we find the lowest
possible bound on the KK scale for the anarchic scenario to be
\be \label{final_anar}
\mkk\gtrsim12 \textrm{ TeV}\,,
\ee
obtained for $Y^*_D=2.0\,$. Interestingly, assuming that the
uncertainty in estimating the mercury EDM in
Eq.~\eqref{anar_hg} is $\sim50\%$, the uncertainty on the
combined bound in Eq.~\eqref{final_anar} is only
$\mathcal{O}(10\%)\,$.

We now switch gears to discuss the flavor triviality case. In
this model, due to the approximate down alignment the dominant
contributions come from the charged Higgs exchange. In the down
mass basis $\left({\rm spurion}\right)_s$ can be written as
\be
\left[D_L f_Q V^{QU} \lambda_U \lambda_U^\dagger V^Q \lambda_D
F_D \right]_{22} \,,
\ee
where $f_X$ and $\lambda_X$ indicate the diagonal forms of
$F_X$ and $Y_X\,$, respectively, while $V^Q$ ($V^{QU}$)
parameterizes the misalignment between $Y_U$ and $Y_D$ ($Y_U$
and $C_Q$) and $D_L$ is the left rotation to the down mass
basis (see Appendix~B in~\cite{Delaunay:2010dw}). Within the RS
linear MFV approximation (where $D_L=\mathds{1}_3$), EDMs are
only induced at the two-loop level~\cite{Fitzpatrick:2007sa,
Banks:1994yg}, similar to the $\theta$-term in the SM. The
leading contribution enters at one-loop order from subleading
terms in the MFV expansion, that are proportional to $\left[Y_U
Y_U^\dagger,Y_D Y_D^\dagger
\right]$~\cite{Kagan:2009bn,Delaunay:2010dw}, and results in a
suppression by $\delta\equiv Y_t^2/Y_b^2$ ($Y_{t,b}$ correspond
to the 5D {\emph{bulk}} top and bottom Yukawas, respectively).
Hence, the dominant contribution to the EDM, which proceeds via
$Y_t$, comes from
\be
\left(D_L\right)_{23} \sim \delta \, V^Q_{23}\,, \quad
V^{QU}_{33} \simeq 1\,, \quad V^Q_{23} \sim r_Q \vckm_{ts} \,,
\ee
where $\vckm$ is the Cabibbo-Kobayashi-Maskawa (CKM) matrix and
$r_Q \equiv f_{Q^3}/ f_{Q^i}$. Another conservative assumption
taken above is in omitting a factor of $Y_b^2$ divided by its
NDA bound, which is necessary for the existence of a phase in
$\left(D_L\right)_{23}\,$. Combining all of the above, we
estimate the $s$ quark EDM:\footnote{We consider only the
contribution from one of the charged Higgs diagrams, as the
other (with the photon attached to the Higgs
line~\cite{Agashe:2008uz}) is of the same order.}
\be \label{ds_edm}
d^E_s \sim \frac{e\,m_{s}}{8 \pi^2 \mkk^2} \, \left( \vckm_{ts}
\right)^2\,Y_t^2\delta \, r_Q^3 \,.
\ee
As a concrete example, we now analyze the sweet spot described
above around Eqs.~\eqref{sweetspot} and~\eqref{sweetspotyuk}.
Plugging these numbers, we find
\be
d_n \sim 4.4 \times 10^{-27} \ecm \simeq 0.15 \, d_n^{\rm
\,exp} \,,
\ee
for $\mkk=1.7$~TeV. In~\cite{Delaunay:2010dw} we reported two
more flavor sweet spots, which include a large 5D bottom Yukawa
and different contributions to CP violation in $B_s$ mixing.
The neutron EDMs for these two sweet spots are roughly 80\% and
110\% of the experimental bound, while other EDMs are much
lower than their corresponding bounds. Given the ${\cal O}(1)$
uncertainties in the associated calculations, both examples can
be considered consistent with present EDM constraints.

\mysection{Collider Phenomenology} The collider phenomenology
of flavor triviality models is interesting, since light
fermions can be composite. This stems from the fact that the EW
fit prefers $c_{U^i}$ to be as composite as possible, although
the $\chi^2$ dependence on this parameter is mild. As a result,
together with a much lower KK scale, hadronic cross sections
are enhanced.\footnote{Throughout this section we set the
KK-gluon coupling to $g_{s*}\simeq 6$ by means of a localized
kinetic term on the UV-brane, which is within the perturbative
regime. As a result of the approximate down alignment, this is
still consistent with flavor constraints.} Specifically, at the
LHC, we expect the KK-gluon production cross section to rise
from the fb regime for anarchic models to the pb one, making it
accessible for early LHC discovery. Furthermore, the
compositeness of the right-handed light quarks potentially
leads to FBAs and to an excess of high-$p_T$ top pairs at the
Tevatron. Below we only focus on existing Tevatron data. We do
not attempt here to provide a complete scan of the parameter
space. Rather, to demonstrate our point that this framework
leads to exciting phenomenology, we evaluate the observables
related to the sweet spot given in Eq.~\eqref{sweetspot}.

The different properties of the KK-gluon compared to the
anarchic scenario warrant a short discussion. First, the
compositeness of some of the light quarks significantly
enhances its production rate, as just mentioned. Conversely,
this also increases $\Gamma_{\rm KK}\,$, the KK-gluon width,
such that\footnote{Decays involving one of the lightest KK
resonances of the custodian fields, which obey $(-,+)$ or
$(+,-)$ boundary conditions and are about 30\% lighter than the
KK-gluon, are suppressed by the EW symmetry breaking scale.
Also, the lightest KK-quarks obeying $(+,+)$ boundary
conditions, those with c=1/2\,, would have the same mass as the
KK-gluon at tree level, while radiative corrections and EW
symmetry breaking effects might render them marginally lighter.
In any case, their effect on the KK-gluon width is
negligible~\cite{Agashe:2003zs, Choi:2002ps, Agashe:2004bm,
Carena:2007tn,Contino:2008hi}.} $\Gamma_{\rm KK} \sim 0.3\,
m_{\rm KK}\,$. However, since we will be interested in energies
which are more than two widths below the mass, it is justified
to ignore effects related to the energy dependence of
$\Gamma_{\rm KK}$ (see \eg~\cite{Barcelo:2011vk} and refs.\
therein for important running width effects for lighter
resonances). Overall it is expected that the prospects for LHC
discovery of the KK-gluon would be greatly increased, for
example via an enhancement of boosted top pair production (see
below). Finally, all of the above implies that this model
should enhance the signal of dijet events, such that it may be
detected or excluded in the future.\footnote{We found our model
to be marginally consistent with recent LHC
data~\cite{Aad:2011aj,Chatrchyan:2011qt}, based on a Monte
Carlo simulation of the rate of central events (rapidity cut
$|y|<0.6$ at the partonic center of mass system) to non-central
events ($|y|<1.7$).}

We now show how this class of models can lead to a large $\ttb$
FBA. Specifically, the asymmetry is enlarged when focusing on
large $\ttb$ invariant masses, as recently found by
CDF~\cite{cdfttbarfbnew}. The asymmetry observed at the
Tevatron reads~\cite{cdfttbarfb,D0ttbarfb,cdfttbarfbnew}
\bea \label{afb_exp}
A^{\ttb}_{450}&\!=&\!(48 \pm 10 \pm 4.9)\%\,,
 \ \, A^{\rm pred}_{450} =(8.8 \pm 1.3)\%
\,, \nonumber \\ A^{\rm lab}_{\rm CDF}&\!=&\!(15 \pm 5.0 \pm
2.4)\%\,, \ \, A^{\rm pred}_{\rm CDF} =(3.8 \pm 0.6)\% \,,
\nonumber \\ A^{\rm lab}_{\rm D0}&\!=&\!(8 \pm 4 \pm 1)\%\,,
 \ \, A^{\rm pred}_{\rm D0} =(1\pm 2)\%\,,
\eea
where $A^{\ttb}_{450}$ is the asymmetry in the $\ttb$ rest
frame for a top pair invariant mass $M_{\ttb}$ larger than
450~GeV, as recently measured by CDF, and $A^{\rm pred}_{\rm
450,CDF,D0}$ is the SM prediction for the corresponding
observable. To make contact with the microscopic new physics
model, it is convenient to replace the lab frame asymmetry with
a $\ttb$ frame one, reported by CDF to be~\cite{cdfttbarfb}
$A^{\ttb}_{\rm CDF}=(16 \pm 7.2 \pm 1.7)\% \,$, while the SM
prediction is $(5.8\pm0.9)\% \,$.

In RS, a differential asymmetry can be generated via $q \bar q$
annihilation into a KK-gluon, which subsequently decays to a
top pair. This requires large axial couplings for \textit{both}
the $q \bar q$ and $t \bar t$ pairs to the KK-gluon, which
arise from large differences between the left and right handed
bulk masses. Since this is a feature of our model (as opposed
to the anarchic case~\cite{Bauer:2010iq}), such an asymmetry is
naturally induced.

\begin{figure}[htb]
\begin{center}
\includegraphics[width=.45\textwidth]{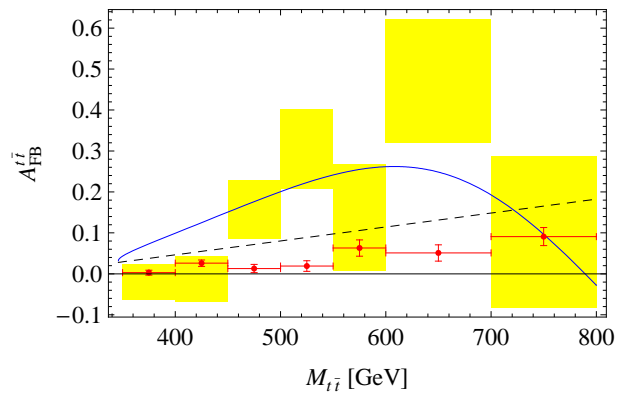}
\caption{Top pair differential forward-backward asymmetry
$A^{t\bar{t}}$ as a function of $M_{\ttb}\,$. Our prediction
(including the SM) is in a solid blue line, while the CDF measurement
(at the detector level) is described by the yellow
shades~\cite{cdfttbarfbnew}. The black dashed line stands for the SM
partonic level prediction computed by MCFM, while the red circles with
error bars correspond to the detector level prediction from
MC@NLO~\cite{cdfttbarfbnew}.}
\label{fig:afb}
\end{center}
\end{figure}

At the partonic level, the asymmetry is given
by~\cite{Djouadi:2009nb,Cao:2010zb}
\be
\hat{A} \propto \beta_t \hat s \left| \mathcal{D} \right|^2
a_t a_q g_{s*}^2 \left[ g_s^2(\hat s-\mkk^2)+2 g_{s*}^2 \hat s
v_t v_q \right]\,,
\ee
where $\mathcal{D}^{-1} \equiv \hat s -\mkk^2 +i \mkk
\Gamma_{\rm KK}$. Here $\hat s$ and $\beta_t \equiv
\sqrt{1-4m_t^2/\hat s}\,$ are the center of mass energy squared
and the top quark velocity, respectively, in the $t \bar t$
frame, and $v_q \equiv -\xi^{-1}+\frac12(f_{q_L}^2+f_{q_R}^2)$
and $a_q \equiv \frac12 (f_{q_L}^2-f_{q_R}^2)$ are the vector
and axial parts of a $q\bar q$ pair to the KK-gluon. We show in
Fig.~\ref{fig:afb} the differential asymmetry as a function of
$M_{\ttb}$ for the above sweet spot parameters\footnote{We
include the SM NLO contribution in a similar way
to~\cite{Cao:2010zb}. We estimate that the uncertainty from the
non-universality of the $k$~factors is $\mathcal{O}$(10\%).},
compared to the recent CDF result~\cite{cdfttbarfbnew}. Note,
however, that the CDF data is not unfolded to the partonic
level, so it cannot be directly compared to the flavor
triviality expectation, yet the overall trend is similar. We
also show NLO Monte Carlo predictions for the SM asymmetry at
the partonic (black dashed curve) and detector (red circles
with error bars) levels. Comparing these two curves, we learn
that the unfolding factor is rather flat. Hence we expect that
the general behavior of the unfolded distribution would be
similar to the CDF one shown in Fig.~\ref{fig:afb}, thus
maintaining the shape agreement between the data and our
prediction.

As an explicit comparison, we note that for the sweet spot of
Eq.~(\ref{sweetspot}) the asymmetry at $M_{\ttb}>450$~GeV is
19\% (including the SM), which is more than $2\sigma$ below the
CDF measurement in Eq.~\eqref{afb_exp}. Yet for the total
asymmetry, our prediction is 12\%, which is less than $1\sigma$
away from the CDF result. At the same time the total $t\bar{t}$
production cross section is $1.2 \sigma$ below the measured
value, while the differential cross section agrees with the CDF
data~\cite{Cao:2010zb}.

Another important consequence of the flavor triviality approach
is an enhanced cross section for the production of high-$p_T$
top pairs, compared to the anarchic RS scenario (although the
branching ratio for the decay of the KK-gluon to top pairs is
smaller by a factor of~$\sim$2). This is particularly
interesting in view of the recent CDF study of boosted massive
jets~\cite{boostedtops, substructure}. This analysis looks for
two massive jets, with mass of $130-210$~GeV and a $p_T$ in the
range of 400-500~GeV. An excess of 3.44$\sigma$ relative to a
simple (yet naive) data driven estimation of the QCD prediction
is observed. If one is to interpret this excess as coming from
new physics, a new source of hadronic tops is required with a
cross section of roughly $11 \pm 3.2$~fb~\cite{gluino}. We find
that our model yields a contribution to the $\ttb$ hadronic
cross of $\sim$5~fb, on top of the SM prediction of
2~fb~\cite{Kidonakis:2003qe, boostedtops}. This is about
1.8$\sigma$ below the observed excess. A possible tension with
the reported measurement is that no excess was found in
hadronic-leptonic top pair events. However, the corresponding
search relies on a large missing energy cut, which tends to be
noisy, with somewhat smaller signal to background
ratio~\cite{boostedtops}. In the case of our prediction above,
this tension is only at the level of 1$\sigma$
(see~\cite{gluino}).

\begin{figure}[htb]
\begin{center}
\includegraphics[width=.47\textwidth]{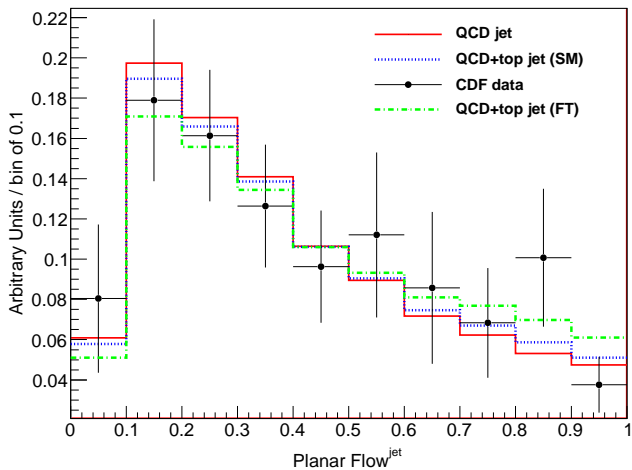}
\caption{Planar flow distribution at the Tevatron, assuming the
following cuts for any given jet: $p_T>400$~GeV, $130<m_{\rm
jet}<210$~GeV, $|\eta|<0.7$, missing $E_T$ significance smaller than
10 and a cone size of 1.0 (anti-kt). The solid red (dotted blue) line
denotes the QCD (QCD+tops) SM prediction, the black circles with error
bars describe the CDF data and the dashed-dotted green line is the
flavor triviality prediction (including the SM).}
\label{fig:pf}
\end{center}
\end{figure}

The excess of top pairs implied above can be detected using jet
substructure analysis techniques. One such example is the jet
shape variable named planar flow (PF)~\cite{topjets} (see
also~\cite{Thaler:2008ju}). High-$p_T$ QCD jets tend to give
low PF values, while top jets lead to higher PF values. In
Fig.~\ref{fig:pf} we present a comparison of the PF
distribution between the SM, our model and the latest CDF
data~\cite{substructure}, for jets with mass of 130-210~GeV and
$p_T$ of 400-500~GeV. We use
MadGraph/MadEvent~\cite{Alwall:2007st} with the Pythia
package~\cite{Sjostrand:2006za} and modified MLM
matching~\cite{Hoche:2006ph}, and the results are interfaced to
FASTJET~\cite{Cacciari:2005hq} for jet clustering. For the SM
QCD + top jet PF distribution, we find a ratio for the SM
$t\bar t$: QCD contributions of 1:13.\footnote{For QCD, we use
MG/ME with a modified MLM matching scheme, while for $t\bar{t}$
events, we rescale the LO MG/ME cross section (without
matching) to the NLO cross section~\cite{Kidonakis:2003qe,
boostedtops}.} This is just to illustrate the method since the
QCD differential cross section has a sizable uncertainty. It is
evident that the RS contribution is somewhat closer to the data
than the pure SM distribution.

\begin{figure}[htb]
\begin{center}
\includegraphics[width=.5\textwidth]{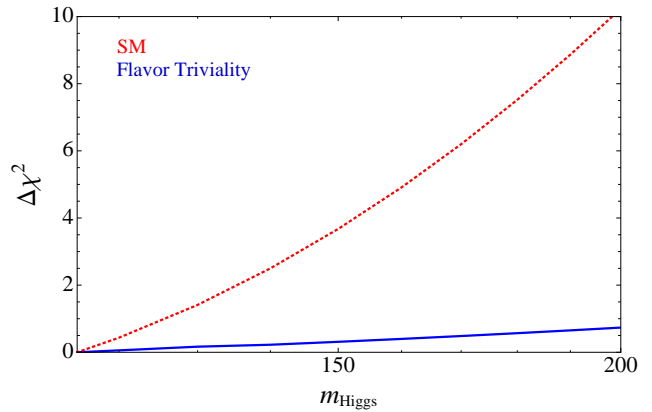}
\caption{Comparison of $\Delta \chi^2$ of the SM and the flavor
triviality model as a function of the Higgs mass.}
\label{fig:higgsmass}
\end{center}
\end{figure}

\mysection{Higgs Mass Dependence} It is known that the
goodness-of-fit of the SM to EW precision observables strongly
depends on the Higgs mass, and rapidly deteriorates when the
latter is raised above the LEP bound. Interestingly, our
model's fit depends only mildly on the Higgs mass, as can be
seen in Fig.~\ref{fig:higgsmass}. Thus, large Higgs mass values
are still compatible with the model, without spoiling the EW
fit (see also~\cite{Agashe:2003zs,Carena:2003fx} for similar
results in RS based on effectively oblique analyses). This is
due to additional contributions to the gauge boson
self-energies, which can be tuned to compensate the SM ones
from a heavier Higgs. In this context, it should be mentioned
that we found another $\chi^2$ minimum for $\mkk\sim$9~TeV,
which is slightly lower than the one reported
in~\cite{Delaunay:2010dw}. However, we choose to cutoff
anything above 4~TeV, hence vetoing excessively fine-tuned
models.

\mysection{Acknowledgments} G.P.\ is the Shlomo and Michla
Tomarin career development chair and supported by the Israel
Science Foundation (grant \#1087/09), EU-FP7 Marie Curie, IRG
fellowship, Minerva and G.I.F., the German-Israeli Foundations,
and the Peter \& Patricia Gruber Award. E.P.\ is supported by
DOE grant DE-FG02-92ER40699.

\end{document}